\documentclass[review]{elsarticle} %twocolumn,3p

\usepackage{lineno,hyperref}
\modulolinenumbers[5]

\journal{Journal of Renewable Energy}
\usepackage{fixltx2e}
\usepackage[utf8]{inputenc}
\usepackage{eurosym}
\usepackage{amsmath}
\usepackage{amssymb}

\usepackage{booktabs}
\usepackage{multirow}

\makeatletter
\def\@author#1{\g@addto@macro\elsauthors{\normalsize%
    \def\baselinestretch{1}%
    \upshape\authorsep#1\unskip\textsuperscript{%
      \ifx\@fnmark\@empty\else\unskip\sep\@fnmark\let\sep=,\fi
      \ifx\@corref\@empty\else\unskip\sep\@corref\let\sep=,\fi
      }%
    \def\authorsep{\unskip,\space}%
    \global\let\@fnmark\@empty
    \global\let\@corref\@empty  %% Added
    \global\let\sep\@empty}%
    \@eadauthor={#1}
}
\makeatother

\begin{document}

\begin{frontmatter}

\title{Assessing the viability of Battery Energy Storage Systems coupled with Photovoltaics under a pure self-consumption scheme}

\author{Georgios~A.~Barzegkar-Ntovom\fnref{a}}
\author{Nikolas~G.~Chatzigeorgiou\fnref{b}}
\author{Angelos~I.~Nousdilis\fnref{c}}
\author{Styliani~A.~Vomva\fnref{c}}
\author{Georgios~C.~Kryonidis\fnref{c}}
\author{Eleftherios~O.~Kontis\fnref{c}}
\author{George~E.~Georghiou\fnref{b}}
\author{Georgios~C.~Christoforidis\fnref{d}\corref{Christoforidis}}
\ead{gchristo@teiwm.gr}
\author{Grigoris~K.~Papagiannis\fnref{c}}

\cortext[Christoforidis]{Corresponding author}

\tnotetext[mytitlenote]{This research has been co-funded by the European Union and National Funds of the participating countries through the Interreg MED Programme, under the project "Promotion of higher penetration of distributed PV through storage for all - StoRES".}

\address[a]{Department of Electrical and Computer Engineering, Democritus University of Thrace, Xanthi, Greece}
\address[b]{FOSS Research Centre for Sustainable Energy / PV Technology Laboratory, Department of Electrical and Computer Engineering, University of Cyprus, Nicosia, Cyprus}
\address[c]{Power Systems Laboratory, School of Electrical and Computer Engineering, Aristotle University of Thessaloniki, Thessaloniki, Greece}
\address[d]{Department of Electrical and Computer Engineering, University of Western Macedonia, Kozani, Greece}

\begin{abstract}
Over the last few decades, there is a constantly increasing deployment of solar photovoltaic (PV) systems both at the commercial and residential building sector. However, the steadily growing PV penetration poses several technical problems to electric power systems, mainly related to power quality issues. To this context, the exploitation of energy storage systems integrated along with PVs could constitute a possible solution. The scope of this paper is to thoroughly evaluate the economic viability of hybrid PV-and-Storage systems at the residential building level under a future pure self-consumption policy that provides no reimbursement for excess PV energy injected to the grid. For this purpose, an indicator referred to as the Levelized Cost of Use is utilized for the assessment of the competitiveness of hybrid PV-and-Storage systems in the energy market, considering various sizes of the hybrid system, battery energy storage costs and prosumer types for six Mediterranean countries. 
\end{abstract}

\begin{keyword}
Battery energy storage systems, grid parity proximity, levelized cost of use, photovoltaics, self-consumption.

\end{keyword}

\end{frontmatter}

\section{Introduction} \label{section:introduction}
Solar photovoltaic (PV) systems have seen a significant growth mainly as a result of the decrease in the prices of the solar PV modules, as well as the reduction of the total installation costs \cite{JRC}. In addition, despite the challenges that the intermittent nature of solar PV systems imposes to the power system, grid-connected residential, i.e. rooftop, PV generation was encouraged by governments worldwide. A range of financial incentives and other supportive schemes have been provided for solar PV uptake, such as various types of Net-Metering schemes or Feed-in Tariffs (FiTs) \cite{Hirth,Riffonneau,RuYu,Darghouth}.
Moreover, in some countries, the reduction in the capital expenses of PV systems and the increased electricity prices resulted to a steadily growing deployment of residential PVs. \par
However, due to the increasing PV penetration in power systems, several technical issues have emerged, mainly related to power quality \cite{KYRITSIS}.
Specifically, the increased PV penetration in the low-voltage (LV) distribution networks, which is the point of connection of the residential PV systems to the electrical grid, may lead to the occurrence of reverse power flow phenomena, possibly resulting in voltage rise above the operational limits, or grid congestion issues \cite{HASHEMINAMIN,WANG}.
PV systems could assist to the reduction of peak load dependence for households; however, this depends to a large extent on both the generation and consumption profiles \cite{Jurasz}. \par
The integration of battery energy storage systems (BESS) alongside PVs at the residential level can mitigate the above mentioned issues. A significant reduction trend in the cost of BESS can be observed in recent years \cite{SolarPower}, assisting to the deployment of BESS along with PV systems as hybrid solutions. As a mismatch between the PV generation and the household consumption profile is usually observed, only a part of the total amount of the household load can be covered by the PV generation directly. Nevertheless, the use of a BESS enables the significant increase of the household’s energy self-consumption and energy self-sufficiency \cite{VIEIRA}, assisting Distribution System Operators (DSOs) at the mitigation of the aforementioned technical challenges \cite{Nousdilis}. Recently, the gradual reduction of FiTs, the steady increase in the electricity prices and other incentives have created a strong motivation for residential PV system owners to increase their PV self-consumption \cite{EU2015}.\par
Given the above and the prospective abolition of FiTs and Net-Metering in the European Union (EU) by 2023, an on-going transition to more cost-oriented approaches, such as self-consumption schemes can be observed \cite{EU}. Such schemes generally favour the exploitation of energy storage systems and thus, there is an increasing interest in the support of storage integration through policy modifications \cite{Mateo}. \par
In this paper, a pure self-consumption policy is considered, i.e. an assumed scheme that provides no reimbursement for excess PV energy injected to the grid. Pure self-consumption is considered for the evaluation of the economic viability of integrated PV and storage systems, since it can be regarded as the most pessimistic scenario, due to the lack of provision of any additional monetary benefits to the prosumer. \par
Among the methodologies provided in the literature for the economic viability assessment of distributed generation projects, an indicator namely the Levelized Cost of Electricity (LCOE) is usually exploited to evaluate the cost of a generating asset \cite{BRUCK}. The LCOE index can be used for the comparison of various electricity generation technologies with different features such as installation capacity, capital cost, lifetime, payback period and risk. Specifically, the LCOE indicator is a way to economically assess the total cost of a power generation system throughout its lifetime, considering the total generated energy over that period. It can also be considered as the minimum cost at which the generated electricity must be sold in order to achieve break-even pricing over the lifetime of a system \cite{Lai}. In general, the main aim of an investment in renewable energy projects and especially of solar PV systems is to reduce the LCOE indicator, rather than just aiming at the reduction of the capital costs \cite{Hernandez}. \par
The well known PV grid parity occurs when the PV LCOE is lower than the retail electricity price. Similarly, the grid parity of a hybrid PV-and-Storage system is reached when the PV-and-Storage system's LCOE is lower than the retail electricity price. However, other parameters must also be considered since BESS operate along with PVs, such as the self-consumption rate (SCR) of the generated energy and the BESS control strategy.  \par
Therefore, a systematic evaluation on how the BESS control strategy affects the LCOE of a hybrid PV-and-Storage system is essential, taking also into consideration the policy under which the hybrid system is operated \cite{BRANKER}. This is important since the different applications offered by the storage asset and thus, the specific BESS control strategy, can influence the lifetime of the battery module \cite{Zakeri}, and subsequently, BESS total cost \cite{JULCH}. For the case of hybrid PV-and-Storage systems the above is considered crucial for their competitiveness on the energy market.
To achieve this, in this paper, a specific index is employed to accurately identify the LCOE of hybrid PV-and-Storage systems, referred to as the Levelized Cost of Use (LCOU). \par
The main contribution of this paper is the evaluation of the competitiveness of hybrid PV-and-Storage systems under a pure self-consumption scheme. For this purpose, the LCOU index is employed to assess whether the grid parity of the hybrid system can be reached under the current market prices or when a future decreased BESS cost is considered. This is of great importance, since the LCOU could also act as an indicator that fewer incentives are required for a specific country towards a future pure self-consumption policy. The LCOU is assumed as a more appropriate indicator given the absence of PV sales, since it does not consider the existence of the power network as a balancing factor compared to the classic definition of the LCOE. \par 
In addition, the paper provides a thorough comparison between the LCOU index of a hybrid system and the retail price when purchasing electricity directly from the grid, considering various sizes of the hybrid system, BESS costs and prosumer types. For this purpose, actual historical data of solar irradiation and load demand for six countries of the Mediterranean region, i.e. Cyprus, Greece, Italy, France, Spain and Portugal, are employed. Within the adopted methodology, the PV generated energy utilized by the households is also taken into consideration. This is performed by calculating the SCR values for a given range of consumption levels in the aforementioned countries derived by both the PV generation and the load demand profiles. \par
\section{Conceptual framework} \label{section:methodology}
The following section reviews the classic definition of the LCOE for a PV system. 
Moreover, it introduces the PV-and-Storage LCOU indicator and verifies its applicability for the assessment of the competitiveness of hybrid PV-and-Storage systems in the energy market. The methodological assumptions and limitations, as well as the categorization of the end-users considered in this work, are also provided.

\subsection{Levelized Cost of Electricity calculation for PV systems}
Equation (\ref{eq:LCOE}) describes the classic definition of the LCOE indicator for a PV system, which determines the production cost of electricity from PVs \cite{tervo}.
\begin{equation}
\label{eq:LCOE}
LCOE_{PV}=\frac{CAPEX_{PV}+\sum_{n=1}^{N}\frac{C_{n}}{(1+r)^{n}}}{\sum_{n=1}^{N}\frac{E_{n}^{produced}}{(1+r)^n}}\\
\end{equation}
In Eq. \eqref{eq:LCOE}, $CAPEX_{PV}$ is the capital expenditure of the PV system, while $C_n$ corresponds to the maintenance costs of year $n$ and $E^{produced}_n$ is the PV generation at year $n$. In addition, {\it N} and {\it r}, is the last year of analysis and the discount factor, respectively.

\subsection{Levelized Cost of Use for hybrid PV-and-Storage systems }
In cases of hybrid PV-and-Storage systems the LCOE of the combined system can also be derived by the standard definition of the PV LCOE, i.e. Eq. \eqref{eq:LCOE}, including the capital and maintenance expenses of both assets.
Moreover, although the $LCOE_{PV}$ constitutes a generalized term, it should be modified when specific policy schemes are examined or when PV is combined with storage. 
In this work, the modified Eq. \eqref{eq:LCOU} is exploited, which is better suited for hybrid PV-and-Storage systems under policy schemes that encourage a higher self-consumption rate of PV generation, where the injected energy to the grid is not reimbursed. Assuming such a policy, the most suitable BESS control strategy would be as follows: charge the battery asset by absorbing the excess power as soon as PV production is greater than consumption, and discharge when PV production is not adequate to supply the load demand \cite{Lai-biogas}.
In Eq. \eqref{eq:LCOU} the use of the self-consumed energy in the denominator can be noticed, resulting to a more representative indicator for the case of PV-and-Storage systems, under the specific BESS operation mode that aims at self-consumption maximization.

\begin{equation}
\label{eq:LCOU}
LCOU=\frac{CAPEX_{PV \& Battery}+\sum_{n=1}^{N}\frac{C_{n}}{(1+r)^{n}}}{\sum_{n=1}^{N}\frac{E_{n}^{produced}\cdot SCR_{n}}{(1+r)^n}}\\
\end{equation}
Note that $SCR_n$ stands for the self-consumption rate of year $n$ as defined in \cite{Cagliari}. It should be clarified that the LCOU term is highly dependent on the $SCR_n$ and consequently on the consumption profile of each prosumer. \par
As already mentioned, the LCOU term is targeted to act as an indicator that fewer incentives are required for a specific country towards a pure self-consumption policy. Specifically, the LCOU indicator does not intend to show the actual economic viability of an investment since other parameters must also be considered, such as the possible provision of any ancillary services. \par
Summarizing, in this work, Eq. \eqref{eq:LCOU} is used to implement the necessary simulations and estimate the corresponding LCOU values for each country, which are then compared with the corresponding retail electricity price. 

\subsection{Methodological assumptions}
Energy storage systems can have many applications, such as power balancing and frequency regulation of power networks. Thus, the LCOE of a PV-and-Storage system differs significantly given the operating conditions of the storage asset and especially of any possible monetary reimbursement of this operation. \par
To this context, the LCOU index was utilized, where a self-consumption maximization mode of the BESS operation without financial benefits for providing any services to the power system is considered. As previously stated, regarding Eq. \eqref{eq:LCOU}, the operation mode of the energy storage asset targets to maximize the self-consumption of the PV generation and does not benefit financially from the interaction with the grid, as there are no PV power sales. Specifically, excess PV generation is stored to the BESS rather than injected to the grid, whereas when PV generation is not adequate to meet the household’s demand the energy stored to the BESS is used. It must be noted that due to this operation mode, only the surplus PV energy can be stored in the BESS. Also, due to the asset’s round-trip efficiency the energy delivered by the energy storage system is reduced \cite{Lai}. Given Eq. \eqref{eq:LCOU}, such losses are taken into consideration through the use of the SCR. This rate is defined as the portion of the PV produced energy, which is finally used for own needs. SCR is enhanced by the use of storage, compared to the case of energy storage absence. Finally, the proposed term is applicable only for newly installed hybrid PV-and-Storage systems, as it incorporates all relevant costs of both PV and BESS in the $CAPEX_{PV\&Battery}$ and $C_n$ terms of the hybrid system, as seen in Eq. \eqref{eq:LCOU}.

\subsection{Methodological limitations}
The SCR of different systems imposes restrictions in the calculation of \eqref{eq:LCOU} as mentioned above, and may result in more system-specific LCOU values. This issue can be resolved by investigating different levels of consumption, in order to determine a more generic case, as presented in subsection \ref{subsection:country analysis}. \par
Furthermore, the LCOU values are highly dependent on the system's location, due to the variation of solar irradiance that has a direct effect on the energy output, and the regional cost differences of the systems \cite{SINGH}. Finally, it should be clarified that the LCOU values are compared with the current electricity prices of each country, without taking into consideration any future increase in the electricity prices and assuming a flat pricing scheme.

\subsection{Categorization of end-users} \label{subsection:country analysis}
The assessment of the economic viability of hybrid PV-and-Storage systems is performed for six countries of the Mediterranean region. In all cases, three different types of consumers are studied so as to generalize the results at a country level. This categorization of end-users is conducted by means of the annual energy demand, as follows:\\
$\bullet$ \textbf{Type A}: 4500~kWh consumption per year.\\
$\bullet$ \textbf{Type B}: 7500~kWh consumption per year.\\
$\bullet$ \textbf{Type C}: 10500~kWh consumption per year.\\
These types are considered to represent low, medium and high end-user consumption, respectively. \par The examined PV system capacity is varied in a predefined range of values, according to the type of prosumers, as determined below:\\
$\bullet$ \textbf{1-5 kWp} PV installed capacity for type A consumers.\\
$\bullet$ \textbf{3-8 kWp} PV installed capacity for type B consumers.\\
$\bullet$ \textbf{5-10 kWp} PV installed capacity for type C consumers.\par
The BESS capacity is determined according to the PV rated power. To examine different BESS capacities the following three ratios are assumed:\\
$\bullet$ \textbf{0.5 kWh}  per kWp. \\
$\bullet$ \textbf{1 kWh}  per kWp. \\
$\bullet$ \textbf{2 kWh}  per kWp. \par

\section{Case studies} \label{section:case studies}
To carry out all the necessary simulations, a tool was developed and then employed for the techno-economic assessment of hybrid PV-and-Storage systems \cite{storestool}.
The tool is capable of giving an insight into the economic viability of hybrid PV-and-Storage systems by providing results of various economic parameters, including the LCOU indicator. A financial analysis for a period of 20 years is performed, taking into consideration the distinct techno-economic characteristics of each country. Note that no battery module replacement is required within the analysis period, since the service life of solar batteries is considered to be greater than 20 years. \par 
\subsection{Typical consumption/generation profiles}
For each country, typical residential generation and consumption profiles are used. Regarding consumption, data sets of typical load profiles are utilized, whereas differences between working and non-working days are also taken into consideration. As far as PV generation profiles are concerned, typical monthly curves are calculated for the capital city of each country, employing the solar radiation database of the PVGIS platform \cite{pvgis}. In Table \ref{table:1}, the annual PV generation in kWh per kWp installed is summarized for all countries. Note that France and Spain present the minimum and maximum PV generation respectively, due to their distinct irradiation characteristics. 

\begin{table}[!b]
\caption{Calculated annual PV generation in kWh per kWp installed, using the solar radiation database of the PVGIS platform.}
\vspace{-2mm}
\label{table:1}
\setlength{\tabcolsep}{2.2em}
\begin{center}
\begin{tabular}{ c | c } 
\hline\hline
\textbf{Country} & \textbf{kWh/kWp}\\
\hline
\(\textbf{Cyprus}\) & 1464.85 \\
\(\textbf{France}\) & 981.08 \\
\(\textbf{Greece}\) & 1368.45 \\
\(\textbf{Italy}\) & 1277.50 \\
\(\textbf{Portugal}\) & 1420.28 \\
\(\textbf{Spain}\) & 1591.61 \\
\hline\hline
\end{tabular}
\end{center}
\end{table}

\subsection{Residential electricity charges and hybrid system cost}
For the calculation of the total electrical energy charges (in \euro/kWh), three parameters are considered, namely the charges for production and supply of electrical energy, charges for electrical networks use, as well as taxes calculated on electrical energy. The total electricity price is obtained by adding the corresponding Value Added Tax (VAT) of each country to the charges. The above-mentioned charges correspond to charges depending on the electrical energy absorbed from the utility grid. However, it should be mentioned that standing fees and charges calculated depending on the installed capacity of the installation are neglected, since they are considered as a fixed paid ammount. In Table \ref{table:2} the total residential electricity cost is summarized for each country. \par
As far as the PV system cost is considered, i.e. PV array and hybrid inverter, a common price of 1300 \euro/kWp is assumed for all countries for ease of comparison \cite{JRC}. Additionally, the corresponding VAT of each country is added to the above mentioned price, whereas it should be mentioned that subsidies are not provided. In terms of the BESS price, two cases are investigated, an assumed current one at 500 \euro/kWh and an expected future one at 150 \euro/kWh. Note that in these values the VAT for each country is also added.

\begin{table}[t]
\caption{Total residential electricity price per country in \euro/kWh, as of April 2019.}
\vspace{-2mm}
\label{table:2}
\setlength{\tabcolsep}{2.2em}
\begin{center}
\begin{tabular}{ c | c  } 
\hline\hline
\textbf{Country} & \textbf{\euro/kWh}\\
\hline
\(\textbf{Cyprus}\) & 0.19270 \\
\(\textbf{France}\) & 0.16814 \\
\(\textbf{Greece}\) & 0.17405 \\
\(\textbf{Italy}\) & 0.21957 \\
\(\textbf{Portugal}\) & 0.20295 \\
\(\textbf{Spain}\) & 0.21780 \\
\hline\hline
\end{tabular}
\end{center}
\end{table}

\section{Results} \label{section:results}
The proposed framework is utilized for the techno-economic evaluation of PV-and-Storage systems, examining various sizes of the hybrid system, BESS costs, and prosumer types, while the corresponding results are presented in this section. Analysis outputs demonstrate the performance of the introduced LCOU term, while also express the PV-and-Storage grid parity proximity for the countries under study.

\begin{figure}[t]
  \centering
  \includegraphics[scale=0.185]{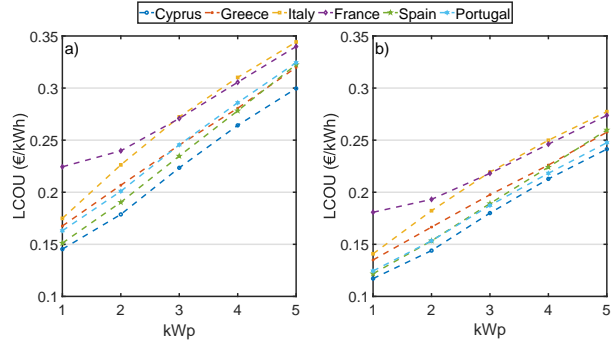} %scale=0.185
  \setlength{\abovecaptionskip}{-1pt}
  \caption{LCOU analysis; 1 kWh/kWp BESS considered. Prosumer type A. Battery module costs: a)~500~\EUR{/kWh}; b)~150~\EUR{/kWh}.}
  \label{fig:FIG-PROSUMER A}
\end{figure}

Analysis outcomes indicate that the LCOU highly depends on the PV-BESS size. Indeed, considering prosumers of type A, the LCOU follows an ascending trend while the system size increases, as shown in Fig.~\ref{fig:FIG-PROSUMER A}a, assuming BESS of 1~kWh/kWp. Moreover, the comparison between the LCOU values and the prices of Table~\ref{table:2} reveals that the operation of a hybrid PV-and-Storage system leads to an LCOU below the retail electricity price for lower PV sizes. Specifically, this can be confirmed for PV systems up to 2~kWp for Cyprus, Portugal and Spain, 1~kWp for Greece and Italy, while in France the LCOU is higher than the current electricity price for residential premises for all cases examined. As expected, a reduction of BESS costs diminishes system LCOU values allowing the investment in larger PV and BESS capacities, as illustrated in Fig.~\ref{fig:FIG-PROSUMER A}b.

\begin{figure}[t]
  \centering
  \includegraphics[scale=0.185]{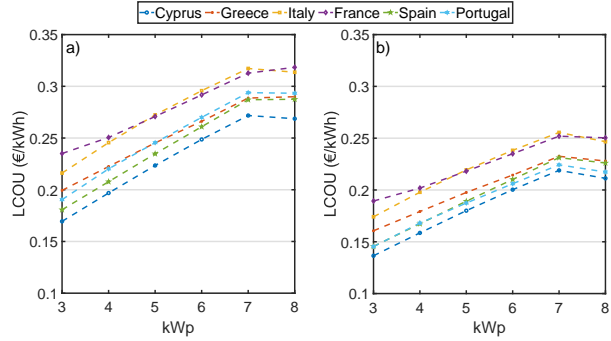} %scale=0.185
  \setlength{\abovecaptionskip}{-1pt}
  \caption{LCOU analysis; 1~kWh/kWp BESS considered. Prosumer type B. Battery module costs: a)~500~\EUR{/kWh}; b)~150~\EUR{/kWh}.}
  \label{fig:FIG-PROSUMER B}
\end{figure}

\begin{figure}[t]
  \centering
  \includegraphics[scale=0.185]{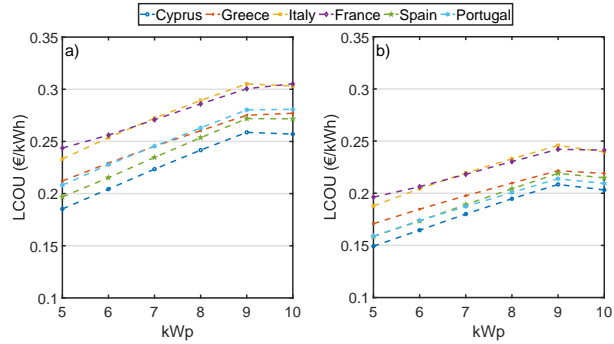} %scale=0.185
  \setlength{\abovecaptionskip}{-1pt}
  \caption{LCOU analysis; 1 kWh/kWp BESS considered. Prosumer type C. Battery module costs: a) 500 \EUR{/kWh}; b) 150 \EUR{/kWh}.}
  \label{fig:FIG-PROSUMER C}
\end{figure}

Analyzing the performance of integrated PV-and-Storage systems owned by prosumers with higher consumption levels, it was revealed that the LCOU increases proportionally to PV-BESS size, however, this trend reverses for prosumers type B and C, when exceeding 7~kWp and 9~kWp PV capacity, respectively, as shown in Figs.~\ref{fig:FIG-PROSUMER B}a and \ref{fig:FIG-PROSUMER C}a. Consequently, it has to be noted that in most cases the LCOU exceeds the current retail price of electrical energy. Nevertheless, this condition changes when decreased BESS costs are considered, as presented in Figs.~\ref{fig:FIG-PROSUMER B}b and \ref{fig:FIG-PROSUMER C}b, making hybrid PV-and-Storage systems profitable for all examined countries excluding France. 

\begin{figure}[t]
  \centering
  \includegraphics[scale=0.175]{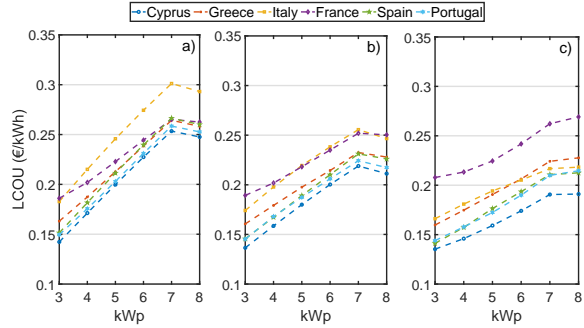} %scale=0.175
  \setlength{\abovecaptionskip}{-1pt}
  \caption{LCOU analysis; Prosumer type B. Battery module costs at 150~\EUR{/kWh}. a)~0.5~kWh/kWp, b)~1~kWh/kWp, c)~2~kWh/kWp.}
  \label{fig:AllBatt_LCOU_TypeB_v2}
\end{figure}

\begin{figure}[t]
  \centering
  \includegraphics[scale=0.19]{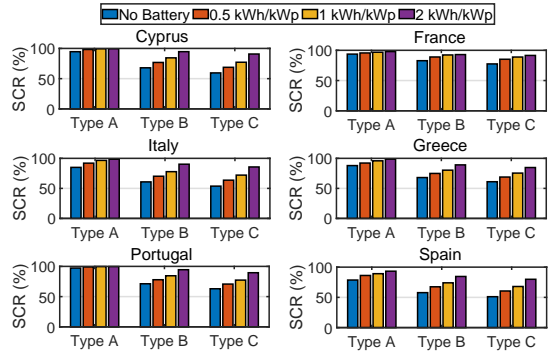} %scale=0.19
  \setlength{\abovecaptionskip}{-1pt}
  \caption{SCR.Type A prosumer: 1 kWp; Type B prosumer: 3 kWp; Type C prosumer: 5 kWp.}
  \label{fig:AllBatt_SCR_TypeABC}
\end{figure}

Furthermore, to assess the impact of BESS capacity on the techno-economic performance of the hybrid systems, the relation between battery energy capacity and PV system rated power is varied as described in Section~\ref{subsection:country analysis}. 
Specifically, Fig.~\ref{fig:AllBatt_LCOU_TypeB_v2} shows that for the same PV size, lower LCOU values can be achieved with larger BESS capacities, considering a battery cost of 150~\EUR/kWh. In fact, BESS operation improves prosumer's SCR and thus the LCOU is reduced. In Fig.~\ref{fig:AllBatt_SCR_TypeABC} the SCR of different prosumer types and various kWh/kWp ratios is illustrated, considering the PV size that provides the lowest LCOU. It should be noted that in contrast to other countries examined, a slight self-consumption improvement is observed by the use of BESS in the case of France, which results in an increase of the LCOU while BESS capacity rises, as depicted in~Fig.~\ref{fig:AllBatt_LCOU_TypeB_v2}.

\begin{figure}[t]
  \centering
  \includegraphics[scale=0.2]{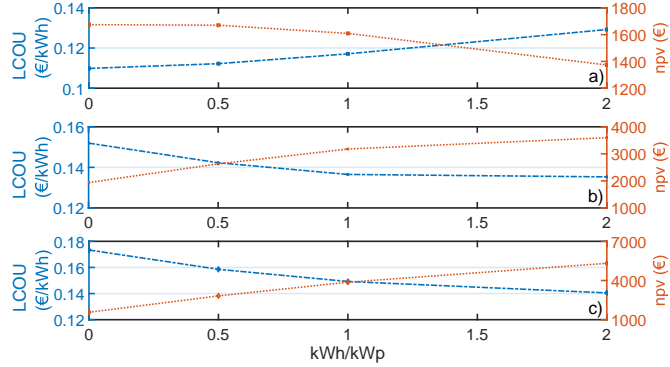} %scale=0.18
  \setlength{\abovecaptionskip}{-1pt}
  \caption{LCOU and NPV analysis per prosumer type considering the PV size that leads to the lowest LCOU; a) Type A: 1~kWp, b) Type B:~3~kWp, c) Type~C:~5~kWp Cyprus.}
  \label{fig:LCOU-NPV_CY}
\end{figure}

Comparative analysis between BESS sizes in Fig.~\ref{fig:LCOU-NPV_CY} demonstrates that integrating BESS with a PV system leads to a lower LCOU for prosumers type B and C in Cyprus, assuming a reduced BESS cost. This is expected since the operation of BESS along with PVs increases the self-consumption of such premises, as shown in Fig.\ \ref{fig:AllBatt_SCR_TypeABC} for the case of Cyprus. In contrast, the SCR of type A prosumers is barely enhanced by BESS, since PV surplus energy is inadequate for charging the battery; consequently the addition of a BESS is not beneficial and may only result in an increase of the LCOU term. The same conclusions would be derived by evaluating the investment's NPV depicted in Fig.~\ref{fig:LCOU-NPV_CY}, confirming the convenience of the LCOU index.

\begin{figure}[b]
  \centering
  \includegraphics[scale=0.19]{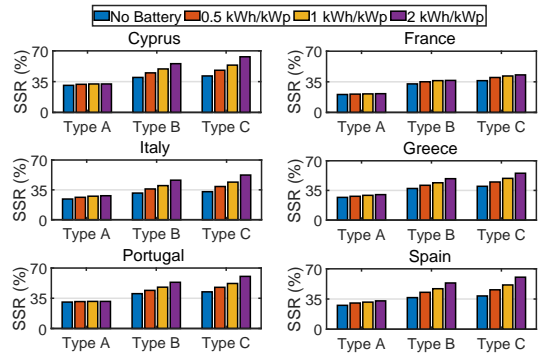} %scale=0.19
  \setlength{\abovecaptionskip}{-1pt}
  \caption{SSR.Type A prosumer: 1 kWp; Type B prosumer: 3 kWp; Type C prosumer: 5 kWp.}
  \label{fig:AllBatt_SSR_TypeABC}
\end{figure}

The previous analyses proved that prosumer's self-consumption plays an important role in the economic evaluation of PV-and-Storage systems, and thus should be taken into account when assessing PV-and-Storage grid parity proximity. The introduced LCOU term incorporates this feature and thus may constitute a useful tool for the feasibility assessment of such hybrid systems, under self-consumption policies with no compensation for surplus energy. Analyses results also prove that when PV-and-Storage grid parity is reached, i.e. the LCOU is lower than the retail electricity price, the NPV of the investment is always positive, thus indicating a viable investment.\par 

\begin{figure*}[t]
  \centering
  \includegraphics[scale=0.270]{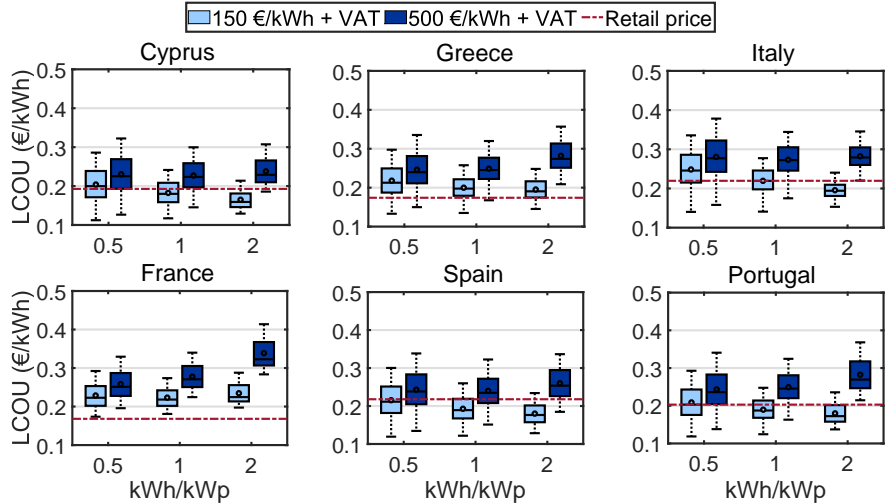} %scale=0.275
  \setlength{\abovecaptionskip}{-1pt}
  \caption{Statistical analysis of LCOU considering all three types of consumers using box plots.}
  \label{fig:boxplot_LCOU_all_prosumers}
\end{figure*}

The SCR and the Self-Sufficiency Rate (SSR), as defined in \cite{Cagliari}, have also been calculated for each case study.  The study confirms that in all countries SCR and SSR of prosumers are augmented by the use of BESS as shown in Figs.~\ref{fig:AllBatt_SCR_TypeABC}~and~\ref{fig:AllBatt_SSR_TypeABC}, especially for premises that present a medium or high consumption level, i.e. prosumers type B and C, respectively. Despite that, the utilization of storage in prosumers of type A slightly improves the performance of the installation when considering a PV capacity of 1~kWp. \par
Finally, the assessment of PV-and-Storage grid parity under a pure self-consumption scheme is conducted for all countries under study utilizing the methodology presented in Section~\ref{section:methodology}, and results are concentrated in the box plots of Fig.~\ref{fig:boxplot_LCOU_all_prosumers}. Specifically, the figure illustrates the allocation of all examined cases per country in terms of LCOU. It can be derived that PV-and-Storage grid parity is rarely reached under the current market prices for storage systems. However, considering future decreased BESS cost, i.e. 150~\EUR/kWh before VAT, the LCOU of all systems under study is generally reduced, while PV-and-Storage grid parity is reached in most cases for Cyprus, Italy, Spain and Portugal. Indicatively, for all the examined BESS sizes, PV-and-Storage grid parity is reached in 61\%, 59\%, 73\% and 63\% of all cases for the above mentioned countries, respectively. In Greece, this only refers to a few individual cases of prosumers, i.e. 22\% of all cases, whereas in France PV-and-Storage grid parity is not reached in any case.

\section{Conclusions} \label{section:conclusions}
In this paper, the LCOU indicator is employed for the competitiveness assessment of hybrid PV-and-Storage systems in the energy market under a pure self-consumption scheme. The LCOU term takes into consideration the self-consumed energy and thus results to a more representative indicator for BESS integrated along with PV systems. The performance of the LCOU is evaluated for six countries of the Mediterranean area considering various sizes of the hybrid PV-and-Storage system, BESS costs and prosumer's types. \par
From the analysis conducted it is shown that the self-consumption rate should be considered when evaluating PV-and-Storage proximity under self-consumption policies with no reimbursement for surplus energy. In addition, it is also derived that in most cases PV-and-Storage grid proximity cannot be reached under the current market prices, unless the cost of BESS is further decreased.

Finally, it has to be recalled that in all cases, similar results were obtained for hybrid system's economic profitability by evaluating the investment's NPV, thus confirming the applicability of the LCOU term on the economic competitiveness assessment of such systems.

%%\section*{References}

%%\bibliography{mybibfile}
%%\bibliographystyle{elsarticle-num}
\bibliographystyle{ieeetr}
\bibliography{mybibfile}{}

\end{document}